\documentclass[11pt,onecolumn,draftcls]{IEEEtran}
\usepackage{amsmath,graphicx,amsfonts,amsthm,bm,textcomp,subcaption,cite}
\usepackage{color}

\newtheoremstyle{mythmstyle}
{3pt}
{3pt}
{\itshape}
{}
{\bfseries}
{:}
{.5em}
{}
\theoremstyle{mythmstyle}

\begin{document}

\title{Permutation Enhanced Parallel Reconstruction with A
Linear Compressive Sampling Device}

\author{Hao~Fang, Sergiy~A.~Vorobyov, and Hai~Jiang%
\thanks{S.~A.~Vorobyov is the corresponding author.
The authors are with the
    Department of Electrical and Computer Engineering, University of Alberta,
    Edmonton, AB, T6G~2V4, Canada; e-mail: \{{\tt hfang2, svorobyo,
    hai1\}@ualberta.ca}.
  }%
}
\maketitle

\begin{abstract}
In this letter, a permutation enhanced parallel reconstruction
architecture for compressive sampling is proposed.
In this architecture, a measurement matrix is constructed
from a block-diagonal sensing matrix and the sparsifying basis of the target signal.
In this way, the projection of the signal onto the sparsifying basis can be
divided into several segments and all segments can be reconstructed in
parallel.
Thus, the computational complexity and the time for reconstruction can be
reduced significantly.
This feature is especially appealing for big data processing.
Furthermore, to reduce the number of measurements needed to achieve
the desired reconstruction error performance, permutation
is introduced for the projection of the signal.
It is shown that the permutation can be performed implicitly by using a
pre-designed measurement matrix.
Thus, the permutation enhanced parallel reconstruction can be achieved with a
linear compressive sampling device.
\end{abstract}

\begin{IEEEkeywords}
Compressive sampling, parallel reconstruction, permutation.
\end{IEEEkeywords}

\section{Introduction}
\label{sec:intro}
In compressive sampling (CS), a signal ${\bf x}$ can be compressed using
a {\it measurement matrix} ${\bf \Phi}$, which results in a
measurement vector ${\bf y} = {\bf \Phi x}$ 
whose length is significantly smaller than that of ${\bf x}$.
If the projection ${\bm \theta}$ of ${\bf x}$ onto an orthonormal
basis ${\bf \Psi}$ is sparse, then ${\bm \theta}$, and thus ${\bf
x}$, can be reconstructed from ${\bf y}$ \cite{Donoho2006},
\cite{Candes2006c}.
This recoverability is guaranteed by the restricted isometry property
(RIP) on the {\it sensing matrix} ${\bf A} \overset{\triangle}= {\bf
\Phi\Psi}$.
Obviously, the measurement vector ${\bf y}$ can be also obtained
from ${\bf A}$ and ${\bm \theta}$ as ${\bf y} = {\bf A}{\bm \theta}$.

It is known that the CS reconstruction process has very high computational
complexity.
Although using a block-diagonal measurement matrix ${\bf \Phi}$ can make
parallel reconstruction of signal ${\bf x}$'s segments possible, the error
performance degrades compared to using a centralized reconstruction scheme
\cite{Gan2007,Gan2008,Duarte2012}.
Recently, a permutation enhanced parallel sampling and parallel
reconstruction architecture has been proposed for CS in
\cite{Fang2012Asilomar,Fang2013TSP}, where segments of sparse projection ${\bm \theta}$ (not
the signal ${\bf x}$) are sampled and reconstructed in parallel.
In \cite{Fang2013TSP}, the sparse projection ${\bm \theta}$ is
reshaped into a 2D matrix and permuted.
All columns of the permuted 2D matrix are then sampled in parallel.
While the reshaping operation leads to parallel reconstruction and the
permutation operation improves the error performance of parallel
reconstruction, the non-linearity introduced by these operations at the
encoder side makes it difficult to implement the sampling device
at a linear system.
To address the above problem, we propose in this letter a way to
preserve the linearity of the encoder. 
It is shown that the permutation enhanced parallel reconstruction can be still
applied without affecting the linearity at the encoder side.

In the proposed architecture, a block-diagonal sensing matrix ${\bf A}$ is
employed, and the measurement matrix ${\bf \Phi}$ for the encoder is
constructed as ${\bf \Phi} = {\bf A} {\bf \Psi}^T$ where $(\cdot)^T$ denotes
the transpose operation.
This is different from the architectures in \cite{Gan2007,Gan2008,Duarte2012}
where a block-diagonal measurement matrix ${\bf \Phi}$ is used and the sensing
matrix ${\bf A}$ for the decoder is constructed based on ${\bf \Phi}$ and
${\bf \Psi}$.
With a block-diagonal sensing matrix, different segments of the projection
${\bm \theta}$ can be reconstructed in parallel, which helps to significantly
reduce the computational complexity and the time needed in reconstruction.
Furthermore, we show that if a permutation that improves the reconstruction
error performance is applied on ${\bm \theta}$, the corresponding measurements
${\bf y}^\dag$ can be obtained by multiplying the signal ${\bf x}$ with
a new measurement matrix ${\bf \Phi}^\dag$.
In this proposed architecture, the projection ${\bm \theta}$ is not required
to be reshaped and explicitly permuted at the encoder side, and thus, it
avoids the non-linearity at the encoder side.
This nice property enables the implementation of the CS sampling
device for the permutation enhanced parallel reconstruction as a linear system
such as the single-pixel camera \cite{Duarte2008}.

\section{Permutation enhanced\\parallel reconstruction for CS}
\label{sec:pcs_rec}
\subsection{Parallel reconstruction}
In this letter, we consider the general case of a multidimensional signal
${\bf x} \in \mathbb{R}^{N_1 \times \dots \times N_d}$.
The vector-reshaped representation of ${\bf x}$ is ${\bf \bar x} \in
\mathbb{R}^{\bar N}$ where $\bar{N} = \prod_{i=1}^d N_i$.
Let the projection of ${\bf \bar x}$ on an orthonormal
sparsifying basis ${\bf \bar \Psi} \in \mathbb{R}^{\bar{N} \times
\bar{N}}$ be $\bm{\bar \theta}$, i.e., ${\bf \bar{x} =
\bar{\Psi}}\bm{\bar \theta}$.
The sparsifying basis $\bar{\bf \Psi}$ can be obtained by the
Kronecker product of several sparsifying bases corresponding to
different dimensions, and thus the orthonormality preserves~\cite{Duarte2012}.
Hence, the measurement vector for ${\bf \bar{x}}$ can be obtained by ${\bf
\bar{y} = \bar{\Phi} \bar{x}=\bar{\Phi} \bar{\Psi}\bm{\bar \theta} =
\bar{A}\bm{\bar \theta}  }$ where ${\bf \bar \Phi}$ is the 
measurement matrix and ${\bf \bar{A} \overset{\triangle}= \bar{\Phi}
\bar{\Psi}}$ is the sensing matrix.
The decoder needs to reconstruct $\bm{\bar \theta}$ from ${\bf \bar{y} =
\bar{A}}\bm{\bar \theta}$.

In the parallel CS reconstruction architecture proposed in this letter, our
objective is to design the sensing matrix ${\bf \bar{A}}$ such that different
segments of $\bm{\bar \theta}$ can be reconstructed in parallel.
We partition $\bm{\bar \theta}$ into $M$ segments such that
\begin{equation*}
\bm{\bar \theta} = [\underbrace{\theta_1 \cdots \theta_{l_1}}_{{\bm
\theta}^T[1]} \underbrace{\theta_{l_1+1} \cdots \theta_{l_1+l_2}}_{{\bm
\theta}^T[2]} \cdots \underbrace{\theta_{{l_1+l_2+\cdots+l_{M-1}+1}} \cdots
\theta_{\bar N}}_{{\bm \theta}^T[M]}]^T
\end{equation*}
where $\theta_i$ denotes the $i$-th element of signal $\bm{\bar \theta}$, and
${\bm \theta}[i]$ denotes the
$i$-th segment with length $l_i$ of signal $\bm{\bar \theta}$.
So we have ${\bar N} = \sum_{i=1}^M l_i$.
Besides, we design the sensing matrix of size $K
\times \bar{N}$ as a block-diagonal matrix, i.e.,
${\bf \bar A} = {\rm diag}( {\bf A}[1], {\bf A}[2], \cdots, {\bf
A}[M])$,
where ${\bf A}[i] \in \mathbb{R}^{K_i \times l_i}$ and $\sum_{i=1}^M K_i = K$.

\begin{figure}[t]
  \centering
    \includegraphics[width=0.65\textwidth]{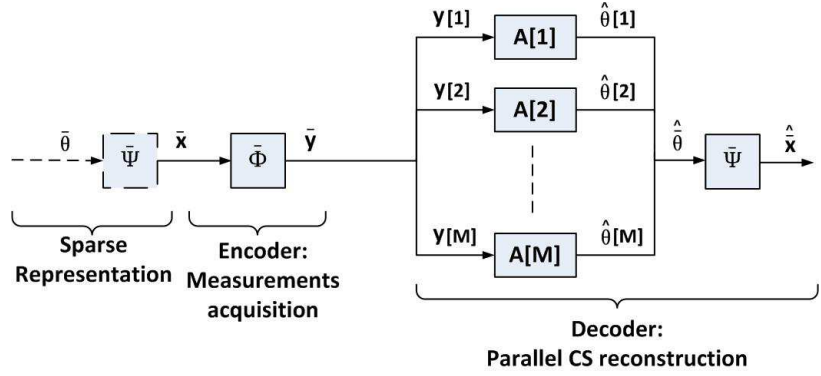}
    \caption{Block diagram of the system employing the parallel CS
      reconstruction
  architecture.}
  \label{fig:block_diag_noperm}
\end{figure}

The block diagram of the system employing the parallel CS reconstruction
architecture is given in \figurename\,\ref{fig:block_diag_noperm}.
In practice, during the acquisition of measurements, $\bm{\bar
\theta}$ is not required to be obtained and ${\bf \bar \Psi}$ is not
required to be stored, which is shown in the diagram by the dashed
line and dashed line border of the corresponding block.
At the encoder side, the measurement matrix is a
pre-designed matrix, which is given by ${\bf \bar{\Phi} = \bar{A}
\bar{\Psi}}^T$, and the measurement vector is given by ${\bf \bar{y} =
\bar{\Phi} \bar{x}}$.
At the decoder side, since ${\bf \bar{A}}$ is block-diagonal,
the measurement vector ${\bf \bar y} =
{\bf \bar{A}}\bm{\bar{\theta}}$ can be divided into $M$ segments, i.e.,
${\bf y}[i] = {\bf A}[i] {\bm \theta}[i]$ ($i=1, 2, \ldots, M$) where ${\bf y}[i]$ denotes the $i$-th
measurement sub-vector.
In this way, all segments of $\bm{\bar \theta}$ can be
reconstructed in parallel, and the signal can be recovered via ${\bf \hat{\bar
x} = \bar{\Psi}}\bm{\hat{\bar \theta}}$ where ${\bf \hat{\bar x}}$ and
$\bm{\hat{\bar \theta}}$ are the reconstructed signal and its projection on
corresponding sparsifying basis, respectively.
In the block diagram shown in \figurename\,\ref{fig:block_diag_noperm},
the block ${\bf A}[i]$ represents the $i$-th CS decoding processor.
The inputs of the CS decoding processors are the
measurement sub-vectors ${\bf y}[i]$'s and the outputs are the reconstructed
segments $\bm{\hat \theta}[i]$'s, which are then stacked in one vector
$\bm{\hat{\bar{\theta}}}$.

Denote the sparsity level of projection $\bm{\bar{\theta}}$ as $S$, i.e.,
there are only $S \ll \bar{N}$ nonzero entries in $\bm{\bar{\theta}}$, and
denote the sparsity level of ${\bm \theta}[i]$ as $S_i$.
Thus, $\sum_{i=1}^M S_i = S$.
It is known that in order to reconstruct exactly ${\bm \theta}[i]$, the sensing
matrix ${\bf A}[i]$ needs to satisfy the RIP condition determined by $S_i$
\cite{Candes2005a}.
Smaller $S_i$ indicates looser RIP condition, and thus $K_i$,
that is, the number
of measurements for ${\bm \theta}[i]$, can be smaller.
Therefore, the design of the sensing matrix ${\bf \bar A}$ depends on
the sparsity levels of the ${\bm \theta}[i]$'s.

\subsection{Computational complexity}
We briefly compare the computational complexity of the parallel CS
reconstruction architecture with that of the architectures employing centralized
reconstruction.
Several solvers exist to reconstruct $\bm{\bar \theta} \in \mathbb{R}^{\bar N}$
from ${\bf \bar{y} = \bar{A} }\bm{\bar{\theta}}$, including for example, the
basis pursuit (BP) algorithm based on interior point methods that have
computational complexity $\mathcal{O}({\bar N}^3)$~\cite{Boyd2004}.
For multidimensional signals, since ${\bar N}$ can be dramatically
large, the reconstruction process becomes rather slow.

For the parallel CS reconstruction architecture in
\figurename\,\ref{fig:block_diag_noperm}, each decoding processor only
needs to reconstruct a segment of the sparse signal $\bm{\bar
\theta}$.
Thus, the computational complexity of the $i$-th decoding processor
is only
$\mathcal{O}(l_i^3)$, and the total computational complexity is
$\mathcal{O}(\sum_{i=1}^M l_i^3)$.
The total computational complexity is minimized to
$\mathcal{O}(\bar{N}^3 / M^2)$
when $l_i = l {=\bar{N}/M}$ for all $i$. 
Thus, by using the parallel CS reconstruction
architecture, the computational complexity is much lower.

\subsection{Permutation}
\label{ssec:perm}
To ensure that all decoding processors in
\figurename\,\ref{fig:block_diag_noperm} have the same configurations,
we assume $l_i = l$ and ${\bf A}[i] = {\bf A}_0$ for all $i$.
Therefore, to reconstruct exactly all ${\bm \theta}[i]$'s, $i=1, 2, \dots, M$,
${\bf A}_0$ needs to satisfy the RIP condition determined by
{$\max_i\left\{S_i\right\}$}.
However, considering the difference of sparsity levels among all ${\bm
\theta}[i]$'s, $i=1, 2, \dots, M$, the above setting is not efficient because fewer
measurements are actually needed for some segments with smaller $S_i$.
This problem can be solved by applying permutation on
${\bm \theta}$ such that all segments have similar sparsity level.
Then, the required number of measurements to achieve a given error
performance can be reduced.

Unlike in \cite{Fang2013TSP} where the permutation
is applied to the 2D-reshaped matrix of ${\bm \theta}$ through an
algorithm-like process, here the permuted projection ${\bm
\theta}^\dag$ is obtained in terms of a linear transform by simply multiplying
${\bm \theta}$ with a permutation matrix ${\bf P}_{\pi}$ defined as follows.
For a permutation $\pi$ of ${\bar N}$ elements: $\{1, \dots, {\bar N}\}
\rightarrow \{1, \dots, {\bar N}\}$ with $\pi(i)$ denoting the
new index of the original $i$-th element after the permutation, its
permutation matrix is ${\bf P}_\pi \in \mathbb{R}^{\bar{N} \times
\bar{N}}$, whose entries are all 0 except that the $\pi(i)$-th entry
in the $i$-th row is 1.

\begin{figure}[t]
  \centering
    \includegraphics[width=0.65\textwidth]{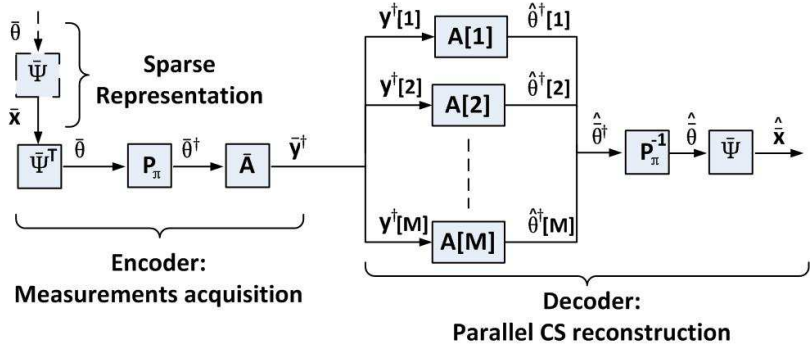}
    \caption{Block diagram of the system employing the parallel CS
      reconstruction architecture with permutation on $\bar{{\bm
      \theta}}$.}
  \label{fig:block_diag_perm}
\end{figure}

\figurename\,\ref{fig:block_diag_perm} describes the parallel CS reconstruction
architecture with permutation on the projection $\bar{\bm \theta}$.
At the encoder side, signal $\bar{\bf x}$ is first projected onto the
sparsifying basis $\bar{\bf \Psi}$, which gives the projection $\bar{\bm \theta}
= \bar{\bf \Psi}^T \bar{\bf x}$.
Then permutation ${\bf P}_\pi$ on the
projection $\bar{\bm \theta}$ is applied, which gives the permuted projection $\bar{\bm
\theta}^\dag = {\bf P}_\pi \bar{\bm \theta}$.
The measurement vector $\bar{\bf y}^\dag$ of permuted projection $\bar{\bm
\theta}^\dag$ is then given using the sensing matrix $\bar{\bf A}$ as $\bar{\bf
y}^\dag = \bar{\bf A}\bar{\bm \theta}^\dag$.
At the decoder side, all segments ${\bm \theta}^\dag[i]$ can be reconstructed in
parallel from ${\bf y}^\dag[i] = {\bf A}[i] {\bm \theta}^\dag[i]$ where
${\bf y}^\dag[i]$ and ${\bm \theta}^\dag[i]$ are the $i$-th
segment of ${\bf \bar y}^\dag$ and the $i$-th segment of $\bar{\bm \theta}^\dag$,
respectively.
The inverse permutation is performed after parallel reconstruction of all
segments of $\bar{\bm \theta}^\dag$, i.e.,
$\bm{\hat{\bar \theta}} = {\bf P}_\pi^{-1} \bm{\hat{\bar
\theta}}{^\dag} = {\bf P}_\pi^T \bm{\hat{\bar \theta}}{^\dag}$.

Actually, the three blocks representing $\bar{\bf \Psi}, {\bf P}_{\pi},
\bar{\bf A}$ can be merged into one, which enables to perform the
measurements acquisition easily in one step as in most CS sampling devices.

Consider a signal ${\bf \bar x} \in \mathbb{R}^{\bar N}$, its
projection $\bar{\bm \theta}$ onto the sparsifying basis $\bar{\bf
\Psi}$, a sensing matrix $\bar{\bf A}$, and a permutation matrix
${\bf P}_\pi$.
The measurements acquired in \figurename\,\ref{fig:block_diag_perm},
i.e., $\bar{\bf y}^\dag = \bar{\bf A} \bar{\bm \theta}^\dag$, can
also be acquired by sampling $\bar{\bf x}$ directly using a
measurement matrix $\bar{\bf \Phi}^\dag$ given by ${\bf
\bar \Phi}^\dag = {\bf \bar{A} P}_\pi {\bf \bar{\Psi}}^T$.
It straightforwardly follows from the associative law, specifically, 
${\bf \bar y}^\dag = \bar{\bf A}\bar{\bm \theta}^\dag 
= {\bf \bar{A} (P}_\pi {\bf \bar \Psi}^T {\bf \bar x}) 
=({\bf \bar{A} P}_\pi {\bf \bar \Psi}^T) {\bf \bar x}$.
Therefore, the same measurements $\bar{\bf y}^\dag$ can be given
using a new measurement matrix
${\bf \bar \Phi}^\dag = {\bf \bar{A} P}_\pi {\bf \bar{\Psi}}^T$.

\begin{figure}[t]
  \centering
    \includegraphics[width=0.30\textwidth]{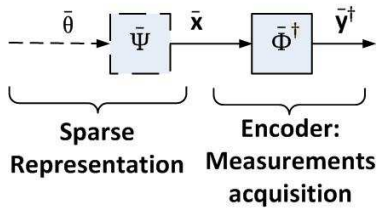}
    \caption{Equivalent encoder to achieve permutation on $\bar{\bm \theta}$.}
  \label{fig:block_diag_perm_acq}
\end{figure}

Accordingly, the encoder in \figurename\,\ref{fig:block_diag_perm}
can be replaced by the encoder shown in
\figurename\,\ref{fig:block_diag_perm_acq}.
Since the new measurement matrix ${\bf \bar \Phi}^\dag$ is pre-generated and
stored at the encoder side, no permutation process is actually required at the
encoder side to obtain ${\bf \bar y}^\dag$.
The only difference between the encoder in
\figurename\,\ref{fig:block_diag_noperm} and that in
\figurename\,\ref{fig:block_diag_perm_acq} is the measurement matrix
$\bar{\bf \Phi} = \bar{\bf A} \bar{\bf \Psi}$ in
\figurename\,\ref{fig:block_diag_noperm} and the new measurement
matrix ${\bf \bar \Phi}^\dag$ in
\figurename\,\ref{fig:block_diag_perm_acq}.

Denote the sparsity level of ${\bm \theta}^\dag [i]$
as $S^\dag_i$.
Recall that we assume $l_i = l$, ${\bf A}[i] = {\bf A}_0$.
If $\max_i\{S^\dag_i\} < \max_i\left\{S_i\right\}$, the
required RIP condition for ${\bf A}_0$ to reconstruct $\bm{\bar
\theta}^\dag$ is weaker than that to reconstruct $\bm{\bar \theta}$.
Thus, with permutation on $\bar{\bm \theta}$, fewer
measurements are needed to achieve the same reconstruction error
performance.
Note that the exact positions of nonzero entries of
$\bar{\bm \theta}$ are not known, and optimal permutation
which results in uniform sparsity levels among all segments of
$\bar{\bm \theta}^\dag$ is not practical.
Thus, permutation design in practice must be based 
on a sparsity model of the projection $\bar{\bm \theta}$, for
example, as introduced in \cite{Fang2013TSP} for video compression
application.
If no sparsity model is known, the best choice of permutation is
a random permutation, which we consider here.
Since we assume $l_i = l$, $\bm{\bar \theta}$ can be rewritten as a
matrix ${\bf \Theta} \in \mathbb{R}^{l \times M}$ by letting ${\bm
\theta}[i]$ be the $i$-th column of ${\bf \Theta}$.
Let $C_i$ be the number of nonzero entries in ${\bf v}_i$, which is the $i$-th
row of ${\bf \Theta}$.
There are $M!$ different permutations that can be applied to ${\bf
v}_i$.
With a random permutation from the $M!$ possibilities, a nonzero entry in ${\bf
v}_i$ is permuted to the columns of $1, 2, \ldots, M$ with equal probability of
$1/M$.
Therefore, considering the permuted $i$-th row, denoted as ${\bf v}^\dag_i$, the
average number of nonzero entries in every column is $C_i/M$.
Assuming that the permutations applied to different
rows are independent from each other, the average number of nonzero entries
in each column of the resulted matrix is $\sum_{i=1}^l C_i/M$.
Therefore, in average, every column of the resulted matrix has the same
sparsity level.
For example, if the projection $\bar{\bm \theta}$ has
length 1000 and we set $l=100$ and $M=10$, for a randomly generated
sparse signal ${\bf \Theta}$ with $S=60$ nonzero entries, after
random permutation, the mean and standard deviation of the sparsity level of
each column of ${\bf \Theta}$ obtained via $10^5$ trials are $6$ and
$2.28$, respectively. 
Therefore, random permutation results in
an acceptable sparsity distribution among segments.

\subsection{Discussion on application}
In most existing CS acquisition devices, the measurement matrix is
pre-generated and stored in the encoder.
The decoder ``stores'' a corresponding sensing matrix for
reconstruction.
The reconstruction in CS is known to have high
computational complexity compared to the sampling process, especially
when the dimension of the signal is very large.
Thus, when the computational complexity and time for
reconstruction are crucial evaluation criteria and centralized
sampling is acceptable, e.g., in real-time video
streaming\cite{Pudlewski2012}, parallel CS
reconstruction is very useful.

The single-pixel camera proposed in \cite{Duarte2008} is
one potential application of the parallel CS
reconstruction architecture.
In the single-pixel camera, the measurements of the image signal
$\bar{\bf x}$ are acquired without digitalization of the analog signal
$\bar{\bf x}$ by high-rate sampling.
To implement the measurement matrix, the single-pixel camera uses a
digital micromirror device (DMD) array with pre-designed random
patterns \cite{Duarte2008}.
However, the DMD array used in the original single-pixel camera can only
represent binary values, whereas in the parallel CS reconstruction
architecture, the entries of the measurement matrix have more than two
values.
This issue can be addressed by using more advanced DMD array.
Actually, contemporary DMD can produce 1024 grayscale value
\cite{Averbunch2012}, and thus, a broader class of measurement
matrices can be represented by such DMD array.

\section{Simulation results}
\label{sec:sim}
We compare the reconstruction error performance and the reconstruction time among
three different schemes: the centralized CS reconstruction, i.e., $M~=~1$; the
parallel CS reconstruction, i.e., $M~\geq~2$, without permutation; and the
parallel CS reconstruction with permutation.
The reconstruction time includes the sum of
reconstruction time of the decoding processors, as well as the average
reconstruction time and the worst reconstruction time of the decoding
processors.

Our simulations are performed using Matlab on a laptop computer with Intel
Dual Core CPU at 2.70~GHz and 8~GB of memory.
The sparse projection $\bar{\bm \theta}$ is a random
binary sequence of length ${\bar N} = 1200$ with $S = 60$ nonzero
entries, which are randomly distributed across the signal.
To ensure that all decoding processors have the same configuration, we set $l_i = l$ and
${\bf A}[i] = {\bf A}_0 \in \mathbb{R}^{K_0 \times l}$ for all $i$, where $K_0=K/M$.
Entries of $\mathbf{A}_0$ are drawn from Gaussian ensembles with variance
$1/K_0$.
The reconstruction algorithm that we use in each decoding processor is the BP
algorithm.
The goal of our simulation is to show the maximum
improvement that can be brought by introducing permutation.
So, ${\bf P}_\pi$ is selected to ensure that all segments of $\bm{\bar
\theta}^\dag$ have the same sparsity level, i.e., $S_i^\dag = S / M$
for all $i$, although such permutation may not be practical.
We set $M$ to 1, 2, 3, 4, 5 and 10.
For $M=1$, the parallel CS reconstruction boils down to the centralized CS
reconstruction.
In our simulation, we run 500 trials
for each combination of ($M, K$) and average the results over the
trails.

\begin{figure}[t]
  \centering
  \begin{subfigure}[b]{0.7\textwidth}
    \centering
    \includegraphics[width=\textwidth]{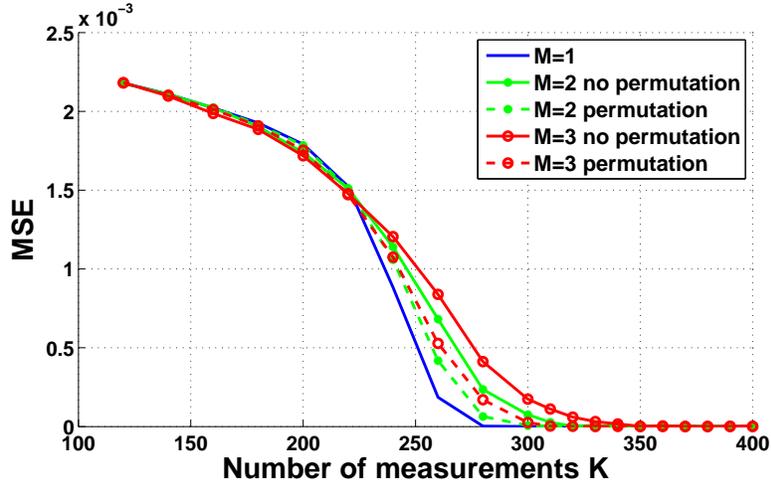}
    \caption{$M~=~1, 2, 3$.}
    \label{fig:mse}
  \end{subfigure}
  \begin{subfigure}[b]{0.7\textwidth}
    \centering
    \includegraphics[width=\textwidth]{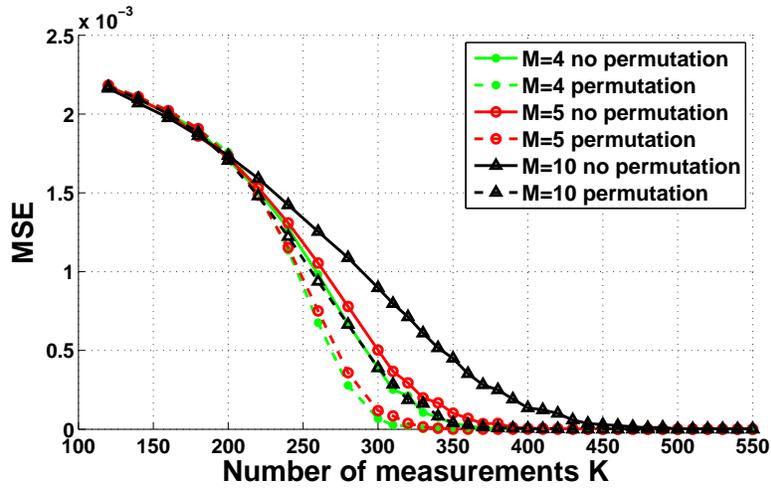}
    \caption{$M~=~4, 5, 10$.}
    \label{fig:mse2}
  \end{subfigure}
  \caption{Reconstruction error performance vs. the number of
  measurements $K$ for different number of segments $M$.}
\end{figure}

\figurename\,\ref{fig:mse} and \figurename\,\ref{fig:mse2} shows the
mean square error (MSE) (normalized to the signal energy)
of the reconstructed signal in  the three aforementioned schemes
versus the number of measurements.
It can be seen that the MSE for a fixed
number of measurements increases as the number of segments $M$
increases.
In other words, the minimum number of measurements required for exact
reconstruction increases as $M$ increases.
It is reasonable since the required number of measurements per segment does not
linearly decrease when $M$ increases.
Besides, it is shown in \figurename\,\ref{fig:mse} and
\figurename\,\ref{fig:mse2} that for a fixed $K$, the MSE can be
reduced with the permutation.
The minimum number of measurements required for exact reconstruction
is also reduced with the permutation.

Table\,\ref{table:rec_time} shows the minimum number of
measurements and the time required for exact
reconstruction.
Here the exact reconstruction is declared if the
normalized MSE of the reconstructed signal is smaller than $2 \times
10^{-5}$.
The total reconstruction time $t_\text{total}$
is the time used to reconstruct all segments.
The average reconstruction time $t_\text{average} =
t_\text{total} / M$ is the average time used to reconstruct each
segment.
The worst reconstruction time $t_\text{worst}$ is the maximal reconstruction
time used to reconstruct the `worst' segment, given as
{$t_\text{worst} = \max_i\left\{t_i\right\}$} where $t_i$ denotes the
reconstruction time for the $i$-th segment.
From Table\,\ref{table:rec_time}, the total
reconstruction time decreases as $M$ increases from 1 to 4. 
When $M$ further increases, the total reconstruction time may increase.
This is because more measurements are required for exact recovery.
However, the total reconstruction time, and the average and the
worst reconstruction time of the parallel CS reconstruction are much less than
those of the centralized CS reconstruction.

\begin{table}[t]
  \centering
    \caption{Minimum number of measurements and reconstruction time
    (in seconds) required for exact reconstruction.}
\label{table:rec_time}
{
\begin{tabular}{|c|c||c||c|c|c|}
\hline
\rule[-1ex]{0pt}{3.5ex} $M$ & Permutation & $K_{\rm require}$ &
$t_\text{total}$ &
$t_\text{average}$ & $t_\text{worst}$\\
\hline
\rule[-1ex]{0pt}{3.5ex} $1$ & N/A & 280 & 4.0243 & 4.0243 & 4.0243 \\
\hline
\rule[-1ex]{0pt}{3.5ex} $2$ & No  & 320 & 1.3946 & 0.6973 & 0.7720 \\
\hline
\rule[-1ex]{0pt}{3.5ex} $2$ & Yes & 300 & 1,3248 & 0.6624 & 0.7290 \\
\hline
\rule[-1ex]{0pt}{3.5ex} $4$ & No  & 370 & 0.8760 & 0.2190 & 0.3041 \\
\hline
\rule[-1ex]{0pt}{3.5ex} $4$ & Yes & 320 & 0.8555 & 0.2139 & 0.3013 \\
\hline
\rule[-1ex]{0pt}{3.5ex} $5$ & No  & 390 & 0.9349 & 0.1870 & 0.2821 \\
\hline
\rule[-1ex]{0pt}{3.5ex} $5$ & Yes & 330 & 0.9132 & 0.1826 & 0.2728 \\
\hline
\rule[-1ex]{0pt}{3.5ex} $10$ & No & 470 & 1.1799 & 0.1180 & 0.1958 \\
\hline
\rule[-1ex]{0pt}{3.5ex} $10$ & Yes & 370 & 1.1943 & 0.1194 & 0.1992 \\
\hline
\end{tabular}
}
\end{table}

\section{Conclusion}
\label{sec:conclusion}
The permutation enhanced parallel reconstruction architecture for
CS has been proposed, where all segments of the projection are reconstructed
in parallel, and the error performance is enhanced by permutation.
It has been shown that via parallel CS reconstruction, the computational
complexity and the reconstruction time can be reduced significantly, with a
certain degree of error performance degradation.
Permutation has been employed to reduce the minimum number of measurements
required for exact reconstruction in the parallel CS reconstruction
architecture.
It has been further demonstrated that permutation can be
implicitly performed via exploiting a new measurement matrix which is the
product of the block diagonal sensing matrix, the permutation matrix, and the
transpose of the sparsifying basis.
Hence, the encoder acts as a linear system, and it enables to use the
proposed permutation enhanced parallel reconstruction with a linear CS sampling
device such as the single-pixel camera.


\end{document}